\documentclass[sigconf]{acmart}

\usepackage{booktabs} 
\usepackage{graphicx}
\usepackage[table,xcdraw]{xcolor}
\usepackage{gensymb}
\usepackage{hyperref}
\usepackage{booktabs}
\usepackage{graphicx}
\usepackage{multirow}
\usepackage{enumitem}
\usepackage{tabularx}
\usepackage{hyphenat}
\usepackage[absolute,overlay]{textpos}

\renewcommand\footnotetextcopyrightpermission[1]{}
\newcommand{\quotes}[1]{``#1''}

\acmConference[ASE-IndustryShowcase '26]{41st IEEE/ACM International Conference on Automated Software Engineering}{October 12--16, 2026}{Munich, Germany}

\begin{document}

\title[Prompt Coach: An Empirical Evaluation of an Agentic Tutor for Learning Prompt Engineering]{Prompt Coach: An Empirical Evaluation of an Agentic Tutor for Learning Prompt Engineering in Software Development}







\author{Rohit Mehra$^\dagger$, Kapil Singi$^\dagger$, Vikrant Kaulgud$^\dagger$, Vibhu Saujanya Sharma$^\dagger$, Swapnajeet Gon Choudhury$^\dagger$, Swati Sharma$^\ddagger$, Adam P. Burden*, Majd Sakr*} 
\affiliation{ 
	\institution{$^\dagger$Accenture Labs, India
		$^\ddagger$Accenture, India
		*Accenture, USA}
	\country{}
}
\email{{{rohit.a.mehra, kapil.singi, vikrant.kaulgud, vibhu.sharma, s.g.choudhury}@accenture.com}}
\email{{{swati.c.sharma, adam.p.burden, majd.sakr}@accenture.com}}

\renewcommand{\shortauthors}{Mehra et al.}

\begin{abstract}

Prompt engineering has emerged as a critical yet undertaught skill for software developers, one that traditional learning approaches are ill-equipped to support given its evolving, interactive, and context-dependent nature. In this paper, we introduce Prompt Coach (PC), an agentic tutor that helps developers learn how to craft high-quality code-generation prompts through Socratic guidance embedded in-flow within their IDE. PC evaluates prompt quality across multiple dimensions and surfaces targeted questions to guide self-correction, grounded in the developer's codebase and the behavior of the target LLM. We present an early empirical study with 15 professional developers combining quantitative prompt quality scoring with qualitative perception measures. Participants showed statistically significant improvements after a single 60-minute session, with the largest gains across dimensions commonly overlooked by developers. They also reported strong trust, high adoption readiness, and unanimous agreement that PC improved their prompt-writing skills.
 
\end{abstract}


\maketitle

\begin{textblock*}{20.5cm}(1.9cm,26.2cm) 
	Accepted for publication at 41st IEEE/ACM International Conference on Automated Software Engineering (ASE 2026 - Industry Showcase)
\end{textblock*}

\section{Introduction}\label{introduction}

With the rapid proliferation of generative and agentic AI, software engineering skills are undergoing a fundamental transformation \cite{app15031344}. Many long-standing skills are becoming obsolete, while a new class of skills is emerging and evolving faster than developers can adapt \cite{10.1145/3696630.3727251, github}. Recent industry reports project that 80\% of developers will need to upgrade their skills by 2027 to remain market relevant \cite{gartner2024upskill}. The focus is shifting from mastering static tools, languages, and frameworks to sustained cognitive collaboration with generative and agentic AI systems. Developers are now required not only to produce correct and efficient code, but also to explicitly articulate intent, constraints, context, and quality expectations through natural language interactions that shape AI-generated artifacts \cite{11334560}. Prompt engineering has consequently emerged as a first-class engineering skill, with organizations increasingly seeking dedicated expertise in this area \cite{10.3389/feduc.2024.1366434, vu2026prompt}. Yet, acquiring such skills efficiently and effectively has become an urgent need, one that traditional learning approaches are ill-equipped to meet.

We posit that, despite their widespread adoption and proven effectiveness in teaching stable programming concepts, traditional learning models (such as video tutorials, books, documentation, online courses, etc.) have fundamental limitations that make them ill-suited for emerging human-AI collaboration skills such as prompt engineering. These limitations include:

\begin{itemize}
	\item \textbf{Lack of context:} Prompt engineering is deeply tied to the developer’s task, codebase, and workflow, whereas videos and tutorials remain generic and detached from real development contexts \cite{drosos2024mytoxictraitthinking}.
	
	\item \textbf{One-size-fits-all calibration:} Developers vary widely in their baseline skills and experience, yet traditional content is typically designed for a single, assumed learner and lacks the ability to adapt dynamically to individual needs \cite{10.1145/3701716.3717527}.
	
	\item \textbf{Inherently interactive practice:} Prompt engineering skill develops through iterative cycles of prompting, observing AI-generated code, and refining prompts, a process that static formats cannot effectively support.
	
	\item \textbf{Rapid content obsolescence:} Static learning materials quickly become outdated as AI systems evolve, often failing to reflect current model behaviors and capabilities \cite{Kuchemann2025}.
	
	\item \textbf{Tacit, experiential expertise:} Much of the expertise involves developing an intuition for model behavior and recognizing subtle failure modes, which is difficult to convey through passive instruction.
	
	\item \textbf{Delayed and generic feedback:} Effective learning requires immediate, context-specific guidance on the prompt being crafted, whereas traditional formats provide delayed and generic feedback \cite{Letourneau2025}.
	
\end{itemize}

As a result of these limitations, learning remains decoupled from day-to-day AI-assisted software development, limiting skill transfer, adaptability, and long-term retention. What is needed is a fundamentally different learning paradigm, one that is adaptive, interactive, embedded within the very workflows in which these skills are exercised, and grounded in the developer's own code and the behavior of the AI systems they target.

We believe that recent advances in agentic AI systems have reached a juncture where they can be meaningfully leveraged to fundamentally transform software engineering education \cite{11201263}. In particular, agentic tutors, defined as LLM-powered systems that can operate autonomously, reason over context, and engage learners through natural, goal-directed interaction, offer a compelling alternative to the traditional learning models discussed above. Unlike static instructional formats, such systems provide adaptive, interactive, and context-aware guidance embedded within the learner’s workflow. Early research has begun to explore this direction for software engineering tasks such as coding and debugging, demonstrating the potential of these systems to improve learning outcomes \cite{kargupta-etal-2024-instruct, 10.1145/3696630.3727255, 11404910}. Importantly, agentic tutors differ from prior AI-based and GenAI-based tutoring systems, which were often limited to static knowledge models or reactive conversational exchanges. They instead combine multi-step reasoning, tool use, and dynamic context adaptation, enabling guidance that is grounded in the learner’s evolving working context.

In this paper, we introduce Prompt Coach (PC), a novel agentic tutor that helps developers learn to craft high-quality code-generation prompts through Socratic guidance. PC is realized as a multi-agent system embedded within the integrated development environment (IDE) that reasons over the developer's prompt in the context of the surrounding codebase and the behavior of the target LLM. It assesses prompt quality across multiple dimensions and surfaces targeted, context-aware questions that guide developers toward iterative refinement through self-correction rather than prescriptive instruction. By design, PC is adaptive, interactive, embedded within the developer's workflow, and grounded in both the developer's code and the behavior of the AI systems they target. To evaluate its impact on learning effectiveness, we present the methodology and results of an empirical study involving 15 professional software developers with diverse experience levels and AI exposure. The study shows measurable improvements in prompt-writing proficiency within a short learning session, along with increased developer confidence and evidence of deeper reflective reasoning about prompt construction.
\section{Related Work}\label{related_work}

Traditional AI-based, and more recently LLM-based, tutoring systems have demonstrated strong learning effectiveness, with LLM tutors outperforming active learning classrooms on outcomes and engagement~\cite{Kestin2025}. Within software engineering, recent works have explored agentic approaches, including guiding students through code debugging~\cite{kargupta-etal-2024-instruct} and simulating multi-agent development teams for learning the software development lifecycle (DevCoach)~\cite{10.1145/3696630.3727255}.

Moreover, several AI-powered tools and resources have recently emerged to support prompt engineering. Anthropic and OpenAI provide interactive tutorials and automated prompt optimizers~\cite{anthropic_tutorial, openai_guide, anthropic_improver, openai_optimizer}. Systematic surveys and pattern catalogs of prompting techniques have been proposed~\cite{schulhoff2024prompt, white2023prompt}, and empirical guidelines exist for code-generation prompts, although a gap persists between guidelines and practice~\cite{midolo2026}. Developers continue to struggle with prompt formulation during exploratory coding interactions~\cite{barke2023grounded}. Tools such as ChainForge enable prompt variation and hypothesis testing across models~\cite{arawjo2024chainforge}, while prompts are treated as structured IDE artifacts without integrated support for improvement~\cite{li2024promptwithme}.

Existing approaches to prompt engineering largely provide static guidance or automated optimization, without adapting to a developer’s task or skill gaps, or supporting the learning process itself. While some tools enable prompt exploration or management, they do not actively teach developers how to construct better prompts. Similarly, existing coaching interfaces offer generic elaboration cues but lack grounding in the developer’s working context and the behavior of the target model. To our knowledge, PC is among the first agentic tutoring systems to address these gaps, delivering adaptive, in-flow guidance grounded in the developer’s codebase and target model behavior, with empirical evidence of measurable improvements in prompt quality.
\section{Prompt Coach}\label{prompt_coach}

\begin{figure}[t]
	\centering
	\includegraphics[width=1.0\linewidth]{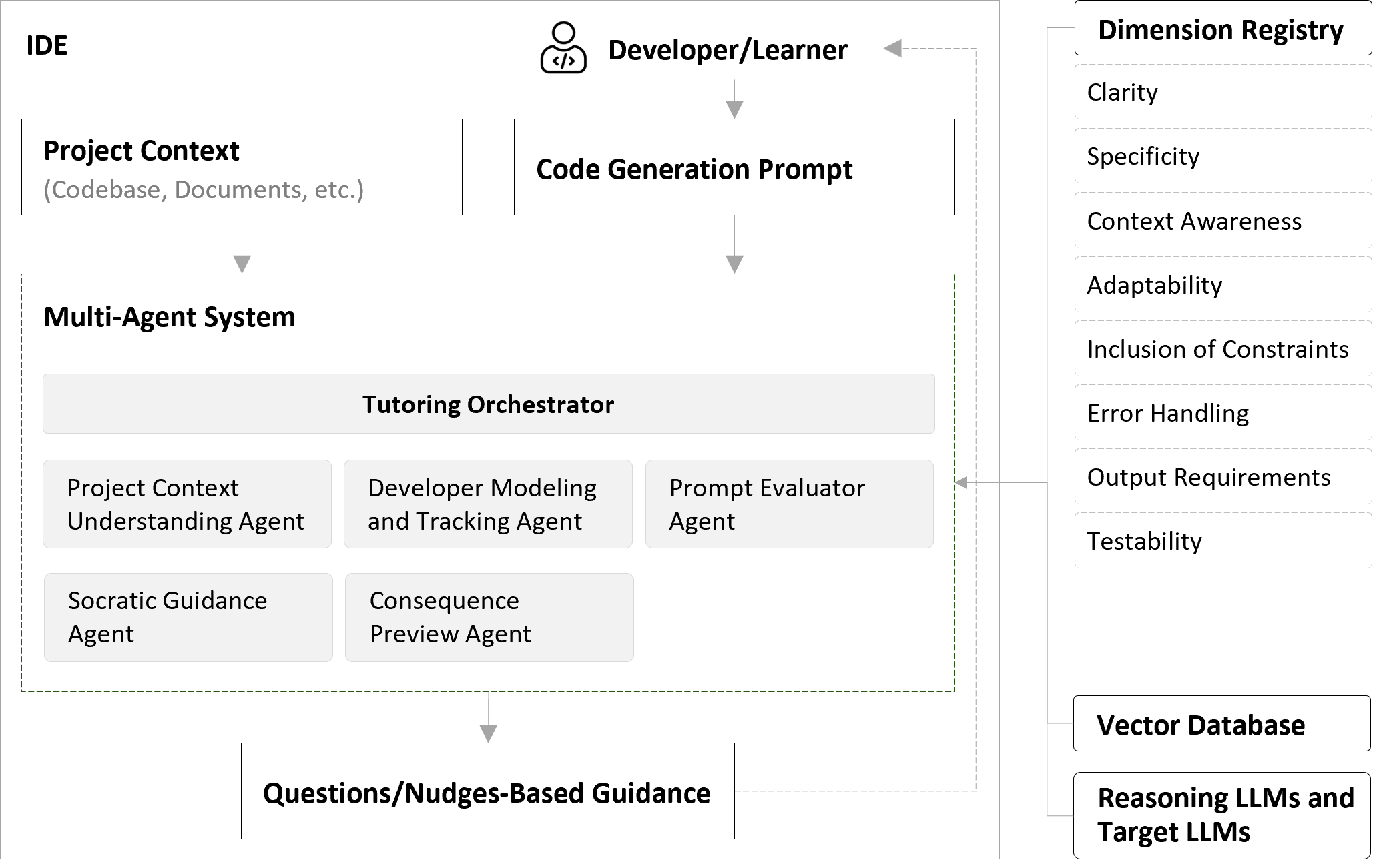}
	\caption{Prompt Coach architecture: An agentic tutor for learning prompt engineering using Socratic guidance.}
	\label{fig:architecture}
\end{figure}

\begin{figure*}[t]
	\centering
	\includegraphics[width=1.0\linewidth]{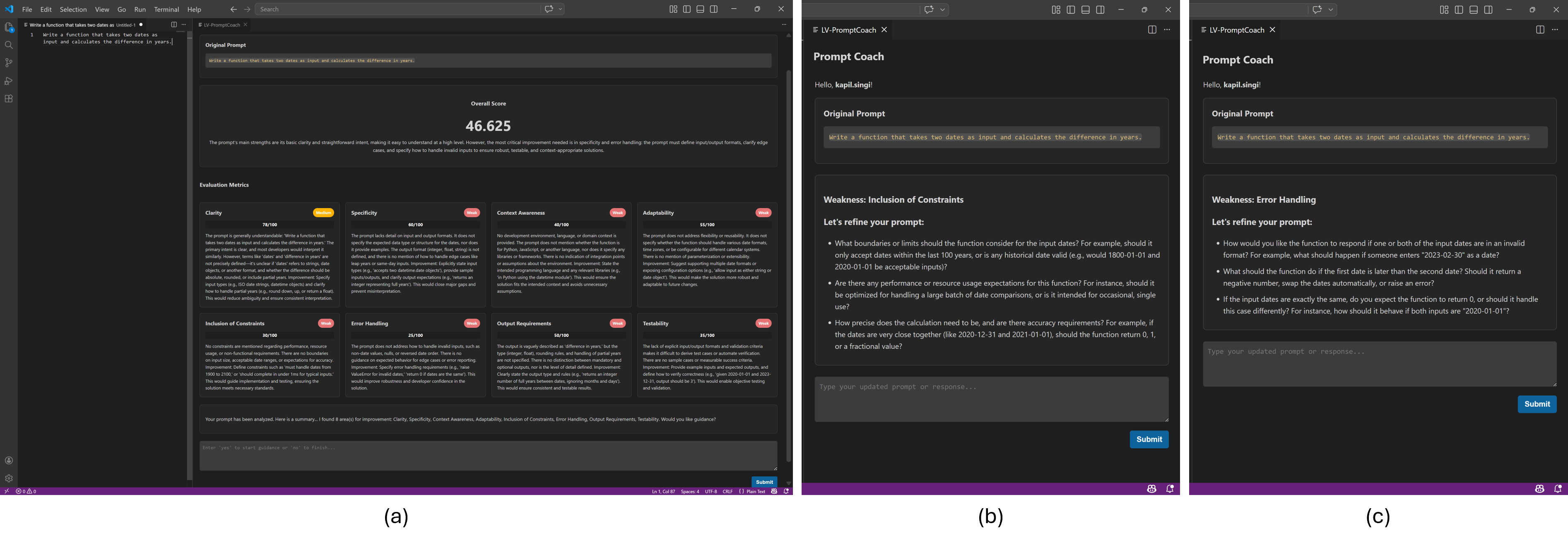}
	\caption{Prompt Coach in action inside the VSCode IDE for an example prompt. (a) Dimensional assessment of the developer's prompt across the eight quality dimensions (b–c) Contextual Socratic guidance surfaced for two weak dimensions, Inclusion of Constraints and Error Handling.}
	\label{fig:screenshot}
\end{figure*}

PC focuses on helping developers learn the skill of crafting high-quality code-generation prompts through Socratic guidance delivered within their IDE. It aims to replace passive, disconnected learning formats with adaptive, in-flow tutoring that responds to the developer's own prompts. Figure \ref{fig:architecture} shows the high-level architecture of our approach.

\subsection{Architecture}

PC is designed as a multi-agent tutoring system that operates directly within the developer's IDE, ensuring that tutoring occurs where code-generation prompts are actually crafted rather than in a separate learning environment. Upon activation, PC ingests the surrounding project context, including the codebase and accompanying artifacts such as design notes, style guides, and requirement specifications (if available). The \textit{Project Context Understanding Agent} then processes this context to extract structural and semantic information, which is embedded into a project-specific knowledge store backed by a vector database. This knowledge store serves as the grounding that downstream agents query for context-aware reasoning. A central \textit{Tutoring Orchestrator} coordinates interactions among agents, routes intermediate outputs, and manages the multi-turn flow between the developer and the system.

Once the project-specific knowledge store is established, PC begins supporting the developer during prompt construction. When the developer submits a code-generation prompt intended for a target LLM, the \textit{Prompt Evaluator Agent} assesses its quality. Rather than applying a monolithic notion of prompt quality, the evaluator scores the prompt across a set of explicitly defined dimensions configured in a \textit{Dimension Registry}. Each dimension receives a score between 0-100, along with a brief rationale, using an LLM-as-a-judge evaluation. For this specific implementation (discussed later in Section 3.2), we adopted the following eight representative dimensions, chosen because they are commonly discussed in the literature as properties of effective code-generation prompts:

\begin{itemize}
	\item \textbf{Clarity:} The prompt clearly communicates intent using well-defined terms and avoids ambiguous language \cite{10.1145/3796535}.
	
	\item \textbf{Specificity:} The prompt explicitly defines inputs, outputs, requirements, and, where applicable, concrete examples \cite{10.1145/3796535}.
	
	\item \textbf{Context Awareness:} The prompt incorporates relevant development context, including environment, dependencies, and intended use case \cite{10.1145/3796535}.
	
	\item \textbf{Adaptability:} The prompt supports variability in inputs and scenarios, and allows for flexible behavior \cite{midolo2026}.
	
	\item \textbf{Inclusion of Constraints:} The prompt specifies technical, performance, resource, sustainability, or security constraints that must be satisfied \cite{10.1145/3722108}.
	
	\item \textbf{Error Handling:} The prompt accounts for edge cases, validation requirements, and expected behavior under failure conditions \cite{10.1109/ASE56229.2023.00143}.
	
	\item \textbf{Output Requirements:} The prompt clearly defines the expected output format, structure, and success criteria \cite{11141320}.
	
	\item \textbf{Testability:} The prompt enables systematic verification of generated code through tests or clear validation criteria \cite{10.1145/3691620.3695527}.
\end{itemize}

Once the prompt has been evaluated, the \textit{Socratic Guidance Agent} transforms the dimensional scores into pedagogical interventions. Rooted in the Socratic method, a dialogue-based approach that guides learners toward insight through targeted questioning rather than direct instruction, this agent generates context-aware nudges that encourage developers to reflect on and refine weaknesses in their prompt instead of prescribing direct fixes \cite{alhossami2024can}. To make this guidance grounded and actionable, the agent draws on two complementary sources of context: the project knowledge store established earlier, which anchors nudges in the developer's own codebase and conventions, and the \textit{Consequence Preview Agent}, which invokes the same class of LLM targeted by the developer to preview the code the current prompt would generate. This preview is not exposed to the developer, as doing so would bypass the learning process by providing a completed artifact rather than encouraging prompt refinement. Instead, it is used internally to identify concrete failure modes arising from the prompt. For example, if a prompt scores low on error handling and the preview indicates that the generated code would fail on empty inputs, the system may ask: \quotes{What should the function do if the input list is empty? Return a default value, raise an exception, or skip silently?} By combining dimensional assessment with project and consequence-grounded reasoning, PC produces guidance that is both targeted to prompt weaknesses and anchored in the developer's working context and the observable behavior of the target LLM, enabling developers to understand not only what to improve, but why those improvements matter.

Finally, as the developer responds to these nudges and refines their prompt, the \textit{Developer Modeling and Tracking Agent} observes each iteration and updates a dynamic representation of the developer’s evolving proficiency in prompt construction. It analyzes how the developer engages with each round of guidance, including which nudges are acted upon, which dimensions improve rapidly, and which remain persistent blind spots, to infer strengths, weaknesses, and areas of growth. This learner profile is continuously updated with each interaction, enabling PC to adapt both the depth and focus of subsequent guidance. For example, if the \textit{Developer Modeling Agent} observes that the developer consistently overlooks error handling across multiple prompts, the \textit{Socratic Guidance Agent} will emphasize questions in that dimension even when subsequent prompts have only minor error-handling gaps. Conversely, dimensions the developer has consistently mastered receive lighter attention, allowing guidance to focus where learning is still needed. By personalizing guidance based on observed behavior rather than a fixed learner profile, PC addresses variability in developer skill levels and supports adaptive, individualized learning. Over time, this enables PC to transition from a static evaluator to a personalized tutor tailored to the developer’s needs.

PC is designed not merely to produce better prompts, but to develop more proficient prompt engineers. By integrating context understanding, adaptive developer modeling, structured prompt evaluation, and Socratic guidance within the IDE, the system turns each code-generation attempt into a reflective learning opportunity. Rather than prescribing what to write, it encourages developers to reason about what is missing, why it matters, and how the surrounding code and target model context shape the outcome. As developers engage with the system over time, this repeated reflection fosters the deeper intuition required for effective prompt engineering, addressing the limitations of traditional learning approaches and enabling in-flow skill acquisition.

\subsection{Implementation}

To demonstrate the applicability of our approach and evaluate its impact on learning, we implemented PC as a working prototype integrated within the VSCode IDE as an extension \cite{vscode}. The system is built as a multi-agent architecture using the CrewAI framework, with backend services deployed on Azure \cite{crewai, azure}. Project context is indexed and retrieved using a ChromaDB-based vector store, enabling context-aware reasoning across agents \cite{chromadb}. Prompt evaluation and guidance are powered by GPT-4.1 from OpenAI, which serves both as the target LLM and as the underlying model for LLM-as-a-judge and consequence preview mechanisms \cite{gpt41}. Figure~\ref{fig:screenshot} presents snapshots of PC in action, illustrating dimensional scoring of prompts and the corresponding contextual Socratic guidance surfaced to the developer.
\section{Empirical Evaluation}\label{empirical_evaluation}

\begin{table}[]
\centering
\ttfamily
\caption{Tasks used in the evaluation study, drawn from the APPS benchmark across three complexity levels.}
\vspace{-5pt}
\label{table:Tasks}
\resizebox{\columnwidth}{!}{%
\begin{tabular}{@{}
l
c
l
>{\raggedright\arraybackslash}p{1.2\linewidth}
@{}}
\toprule
\textbf{Phase}                          & \multicolumn{1}{l}{\textbf{APPS ID}} & \textbf{Complexity} & \textbf{AI-Refined Description}                                                                                                                                                                                                                                                                                                                                        \\ \midrule
\multirow[t]{3}{*}{\textbf{Baselining}}    & 4987                                      & Introductory        & Write a function that takes two dates as input and calculates the difference in years.                                                                                                                                                                                                                                                                                 \\
                                        & 1961                                      & Interview           & Create a program that simulates a browser with basic history navigation. It should allow starting from a homepage, visiting new pages, and moving backward or forward through the visited history.                                                                                                                                                                    \\
                                        & 2207                                      & Competition         & Write a program that simulates pings to two servers. Each ping sends multiple packets, some of which succeed and some fail. Based on the ping results, determine whether each server should be considered “alive” (at least half of its packets succeeded) or “dead”.                                                                                                  \\ \midrule
\multirow[t]{3}{*}{\textbf{Post-Learning}} & 2815                                      & Introductory        & Write a program that takes a sentence as input, builds a list of its words, and outputs a string representing the positions of each word in that list.                                                                                                                                                                                                                 \\
                                        & 1924                                      & Interview           & Write a program that checks a list of transactions and flags ones that break certain validity rules, such as large amounts or conflicts with other transactions.                                                                                                                                                                                                       \\
                                        & 2016                                      & Competition         & Write a program to help Alex plan a trip across cities connected by roads. Each city has an associated fun score, and Alex wants to maximize the total fun score of the cities he visits. He starts in one city, can revisit cities, but may not immediately travel back on the same road he just used. The goal is to find the maximum possible score he can achieve. \\ \bottomrule
\end{tabular}%
}
\end{table}

We conducted an early empirical evaluation to assess the impact of PC on developers' prompt-engineering proficiency, using a mixed-methods design that combined quantitative measurement of prompt-writing proficiency with qualitative assessment of developer perceptions. The study was guided by the following research questions:

\textit{\begin{itemize}
	\item \textbf{RQ1:} What is the baseline quality of prompts that developers write when performing code-generation tasks?
	\item \textbf{RQ2:} What is the measured learning impact of PC on developers' prompt-engineering proficiency?
	\item \textbf{RQ3:} How do developers perceive and reflect on using PC as an agentic tutor embedded in their daily coding workflows?
\end{itemize}}

\subsection{Methodology}

\begin{table}[]
	\centering
	\ttfamily
	\caption{Baseline and post-learning prompt-quality scores, with relative improvement and paired Wilcoxon signed-rank {$\mathit{p}-values$}.}
	\vspace{-5pt}
	\label{table:Quantitative}
	\resizebox{\columnwidth}{!}{%
		\begin{tabular}{@{}lccc>{\centering\arraybackslash}p{1.0cm}@{}}
			\toprule
			& \multicolumn{1}{l}{\textbf{Avg. Baseline}} & \multicolumn{1}{l}{\textbf{Avg. Post-Learning}} & \multicolumn{1}{l}{\textbf{\%Increase}} & \multicolumn{1}{l}{\textbf{$\mathit{p}-value$}}   \\ \midrule
			\textbf{Overall}         & 63.04                                      & 71.69                                           & + 13.73 \%                              & \cellcolor[HTML]{EFEFEF}\textless .001  \\ \midrule
			\textbf{Complexity-Wise} &                                            &                                                 &                                         &                                         \\
			Introductory             & 65.63                                      & 74.48                                           & + 13.93 \%                              & \cellcolor[HTML]{EFEFEF}0.003           \\
			Interview                & 62.56                                      & 69.05                                           & + 10.38 \%                              & \cellcolor[HTML]{EFEFEF}0.004           \\
			Competition               & 60.66                                      & 71.34                                           & + 17.71 \%                              & \cellcolor[HTML]{EFEFEF}0.001           \\ \midrule
			\textbf{Dimension-Wise}  &                                            &                                                 &                                         &                                         \\
			Clarity                  & 79.87                                      & 79.71                                           & - 00.19\%                                & 0.577                                   \\
			Specificity              & 68.58                                      & 72.07                                           & + 05.09\%                                & \cellcolor[HTML]{EFEFEF}0.003           \\
			Context Awareness        & 56.56                                      & 69.91                                           & + 23.61\%                               & \cellcolor[HTML]{EFEFEF}0.002           \\
			Adaptability             & 66.89                                      & 69.58                                           & + 4.02 \%                               & 0.208                                   \\
			Inclusion of Constraints & 50.51                                      & 66.49                                           & + 31.63\%                               & \cellcolor[HTML]{EFEFEF}\textless .001  \\
			Error Handling           & 52.56                                      & 68.67                                           & + 30.66\%                               & \cellcolor[HTML]{EFEFEF}\textless{}.001 \\
			Output Requirements      & 70.07                                      & 77.91                                           & + 11.20\%                               & \cellcolor[HTML]{EFEFEF}\textless .001  \\
			Testability              & 59.29                                      & 69.20                                           & + 16.72\%                               & \cellcolor[HTML]{EFEFEF}0.001           \\ \bottomrule
		\end{tabular}%
	}
\end{table}

We recruited 15 professional software developers from our delivery centers to participate in the study. The participants recorded a mean development experience of 9.6 years (ranging from 3 years to 22 years). Participants were selected only if they used AI-coding assistants in their day-to-day roles, regardless of whether they had previously taken prompt-engineering training.

The study followed a single-arm, within-subjects, pre/post design conducted remotely via video conferencing, with a moderator guiding each session. This design was chosen to establish an initial within-subject measurement of PC's learning impact before extending to comparative evaluations against alternative learning approaches in future work. The study was structured across five phases, with an average duration of approximately 2 hours and 30 minutes per participant:

\begin{itemize}
	
	\item \textbf{Pre-Study Questionnaire:} Participants completed a questionnaire capturing demographic and background information, including development experience, career level, prior prompt\hyp{}engineering training, and frequency of AI coding assistant usage.
	
\begin{table*}[t]
	\ttfamily
	\centering
	\caption{Post-study qualitative questionnaire results captured on a 7-point Likert scale and one-sided Wilcoxon signed-rank {$\mathit{p}-values$} tested against the neutral midpoint. (\% Agreement = participants responding $\geq 5$, and IQR = Interquartile range)}
	\vspace{-5pt}
	\label{table:Qualitative}
	\resizebox{\linewidth}{!}{%
	\begin{tabular}{
			@{}
			>{\raggedright\arraybackslash}p{0.13\linewidth}
			>{\raggedright\arraybackslash}p{0.3\linewidth}
			l
			>{\raggedright\arraybackslash}p{0.7\linewidth}
			c
			c
			c
			c
			c
			>{\centering\arraybackslash}p{1.1cm}
			@{}
		}
		\toprule
		\textbf{Category}                                   & \textbf{Definition}                                                                                   & \textbf{Parameter}                  & \textbf{Question asked to the participant}                                                                                                                                                   & \textbf{Mean} & \textbf{SD} & \textbf{Median} & \textbf{IQR} & \textbf{\% Agreement} & \textbf{$\mathit{p}-value$} \\ \hline \noalign{\vskip 2pt}
		\multirow[t]{5}{*}{\parbox[t]{1.0\linewidth}{\raggedright\arraybackslash \textbf{Learning Effectiveness}}}    & \multirow[t]{5}{*}{\parbox[t]{1.0\linewidth}{\raggedright\arraybackslash Perceived improvement in prompt-writing skills and quality}}    & \textbf{Learning Intervention}      & The learnings were effective in improving my prompt-writing skills.                                                                                                                          & 6.33          & 0.62        & 6.0             & 1.00         & 100\%                & \cellcolor[HTML]{EFEFEF}\textless{}0.001 \\
		&                                                                                                       & \textbf{Prompt Quality}             & I believe that the prompts I created post-learning could lead to higher quality generated code.                                                                                              & 5.73          & 1.79        & 6.0             & 1.00         & 86.7\%               & \cellcolor[HTML]{EFEFEF}0.013            \\
		&                                                                                                       & \textbf{Prompting Velocity}         & Post-learning, I spent more time in crafting high-quality code-generation prompts.                                                                                                           & 5.73          & 0.96        & 6.0             & 0.50         & 93.3\%               & \cellcolor[HTML]{EFEFEF}0.002            \\ \hline \noalign{\vskip 2pt}
		\multirow[t]{3}{*}{\parbox[t]{1.0\linewidth}{\raggedright\arraybackslash \textbf{Cognitive Depth}}}           & \multirow[t]{3}{*}{\parbox[t]{1.0\linewidth}{\raggedright\arraybackslash Perceived enhancement in thinking, reasoning, and self-reflection during prompt writing}} & \textbf{Cognitive Elaboration}      & Post-learning, I have started thinking more about what other details should be specified in my prompt.                                                                                       & 6.20          & 1.57        & 7.0             & 1.00         & 93.3\%               & \cellcolor[HTML]{EFEFEF}0.010            \\
		&                                                                                                       & \textbf{Confidence}                 & The learnings improved my confidence to write effective code-generation prompts.                                                                                                             & 5.93          & 1.39        & 6.0             & 1.00         & 86.7\%               & \cellcolor[HTML]{EFEFEF}0.005            \\ \hline \noalign{\vskip 2pt}
		\multirow[t]{2}{*}{\parbox[t]{1.0\linewidth}{\raggedright\arraybackslash \textbf{Trust and Personalization}}} & \multirow[t]{2}{*}{\parbox[t]{1.0\linewidth}{\raggedright\arraybackslash Perceived trust in the tutor and personalization of its guidance.}}      & \textbf{Trust}          & I trust these learnings to guide me in crafting better code-generation prompts.                                                                                                              & 6.00          & 1.07        & 6.0             & 1.00         & 93.3\%               & \cellcolor[HTML]{EFEFEF}0.001            \\
		&                                                                                                       & \textbf{Perceived Personalization}  & Learnings were contextual/personalized based on my inputs.                                                                                                                                   & 5.73          & 1.16        & 6.0             & 1.50         & 86.7\%               & \cellcolor[HTML]{EFEFEF}0.003            \\ \hline \noalign{\vskip 2pt}
		\multirow[t]{6}{*}{\parbox[t]{1.0\linewidth}{\raggedright\arraybackslash \textbf{Adoption Readiness}}}        & \multirow[t]{6}{*}{\parbox[t]{1.0\linewidth}{\raggedright\arraybackslash Likelihood of adopting, recommending, and continuing to use Prompt Coach}}      & \textbf{Adoption Likelihood}        & I would consider integrating Prompt Coach into my regular development workflow for continuous learning.                                                                                      & 5.73          & 1.39        & 6.0             & 2.00         & 86.7\%               & \cellcolor[HTML]{EFEFEF}0.009            \\
		&                                                                                                       & \textbf{Recommendation Likelihood}  & I would recommend Prompt Coach to other developers within my team for improving their prompt-writing skills.                                                                       & 6.00          & 1.31        & 6.0             & 1.00         & 93.3\%               & \cellcolor[HTML]{EFEFEF}0.004            \\
		&                                                                                                       & \textbf{Engagement}                 & Prompt Coach's Socratic guidance approach was engaging.                                                                                                         & 5.67          & 1.63        & 6.0             & 1.50         & 80.0\%               & \cellcolor[HTML]{EFEFEF}0.013            \\
		&                                                                                                       & \textbf{Comparative Perception}     & Compared to other learning approaches (like books, blogs, articles, tutorials, and MOOCs), I perceive Prompt Coach as more useful for improving my prompt-writing skills. & 5.80          & 1.47        & 6.0             & 1.50         & 80.0\%               & \cellcolor[HTML]{EFEFEF}0.009            \\ \bottomrule
		\end{tabular}%
}
\end{table*}
	
	\item \textbf{Baselining Phase:} Participants wrote code-generation prompts for three tasks spanning introductory, interview, and competition-level complexity, sampled from the widely used APPS (Automated Programming Progress Standard) benchmark \cite{NEURIPS} (10 minutes per task). To prevent verbose task descriptions from artificially inflating prompt quality scores, each description was refined using an LLM to remove extraneous details while preserving task intent. Table \ref{table:Tasks} lists the tasks used in the study. Participants had no prior exposure to PC at this stage.
	
	\item \textbf{Learning Phase:} Participants engaged in self-directed learning using PC with the three prompts crafted in the previous phase (60 minutes). The moderator first conducted an end-to-end demonstration of PC and assisted participants in setting it up on their systems before the learning phase began.
	
	\item \textbf{Post-Learning Phase:} Participants wrote code-generation \\prompts for three additional APPS tasks at the same complexity levels (10 minutes per task), with access to PC revoked during this phase. The tasks were distinct from those in the baselining phase to prevent memorization effects.
	
	\item \textbf{Post-Study Questionnaire:} Finally, participants completed a questionnaire capturing their perceptions of PC and the learning experience. All responses were recorded on a 7-point Likert scale (1 = Strongly Disagree, 7 = Strongly Agree).

\end{itemize}

Prompts collected during the baselining and post-learning phases were subsequently scored across the eight quality dimensions defined in Section~\ref{prompt_coach}, using an LLM-as-judge evaluation with the same model and prompting setup as the \textit{Prompt Evaluator Agent}. The resulting per-participant, per-dimension prompt-quality scores form the basis for the quantitative analysis addressing RQ1 and RQ2, while responses from the post-study questionnaire inform RQ3.

\subsection{Results}

\textbf{RQ1: Baseline quality of developer-written prompts}: Across the three baselining tasks, participants achieved a mean prompt-quality score of 63.04, indicating that professional software developers produce prompts of moderate quality. Table \ref{table:Quantitative} summarizes the quantitative results. Further, baselines were broadly consistent across complexity levels (introductory: 65.63, interview: 62.56, competition: 60.66). A Pearson correlation ($r = 0.11$) between years of experience and baseline score indicates that prompt-writing proficiency does not meaningfully scale with programming experience, suggesting it is a distinct skill and an emerging area of expertise. Dimension-wise, however, variation was striking: developers scored highest on Clarity (79.87) and Output Requirements (70.07), but lower on Inclusion of Constraints (50.51), Error Handling (52.56), and Context Awareness (56.56). This pattern suggests that while developers naturally produce clear prompts that communicate intent, they tend to underspecify constraints, edge cases, and contextual details critical for downstream code generation.

\textbf{RQ2: Learning impact of PC}: After a 60-minute learning session with PC, post-learning prompts achieved a mean score of 71.69, representing a 13.73\% relative improvement over the baseline (ranging from -0.28\% to 38.37\%). Notably, 13 out of 15 participants showed improvements, while 2 participants exhibited no measurable gain on average. Importantly, no participant showed a meaningful decline in performance. The slight negative change (-0.28\%) can be attributed to minor variability inherent in LLM-as-a-judge evaluations, which may introduce small fluctuations in scores across runs. Statistical significance was assessed using the paired Wilcoxon signed-rank test, with significant results ($p < 0.05$) highlighted in grey (ref. Table \ref{table:Quantitative}). These results highlight the positive impact of PC on improving prompt-writing proficiency within a relatively short learning period. Notably, these gains were achieved after a single 60-minute session, suggesting that sustained interaction with the system may yield further improvements. Furthermore, the largest gains were observed in Inclusion of Constraints, Error Handling, and Context Awareness, which were also the lowest-scoring dimensions at baseline, suggesting that PC is particularly effective at addressing developers' cognitive blind spots.

\textbf{RQ3: Learner perceptions of PC}: Participants reported strong positive sentiment across all learning dimensions. Table \ref{table:Qualitative} summarizes the results. Statistical significance was assessed using a one-sample Wilcoxon signed-rank test against the neutral midpoint, with Holm-Bonferroni correction, with significant results ($p < 0.05$) highlighted in grey. All parameters exhibited mean scores ranging from 5.67 to 6.33 on a 7-point Likert scale. Participants unanimously agreed that PC improved their code-generation prompt-writing skills (100\% agreement), expressed strong trust in the guidance provided, and reported increased cognitive elaboration during prompt construction. Adoption readiness was similarly high, with most participants indicating that they would integrate PC into their regular workflow and recommend it to colleagues.

\subsection{Threats to Validity}

Our findings are based on 15 developers from a single organization using a single code-generation benchmark (APPS), which may limit generalizability. To keep the study mechanics consistent across participants, the study was conducted in isolation without access to real project context, which PC is designed to leverage during actual development workflows. Finally, due to limited participant availability, the learning session was restricted to 60 minutes. Longer engagement may yield different patterns of impact. Despite these considerations, the large observed effect size and participant-level consistency support the robustness of our findings.
\section{Conclusion and Future Work}\label{conclusion}

This paper presents our work on leveraging agentic AI to transform how developers learn emerging skills such as prompt engineering. We introduced PC, an agentic tutor that delivers adaptive, contextual, Socratic guidance within the IDE to help developers learn how to craft high-quality code-generation prompts. An initial empirical evaluation with 15 professional developers demonstrated statistically significant improvements in prompt-writing proficiency after a single 60-minute session. Participants also reported strong positive perceptions of the learning experience. These findings position agentic tutoring as a promising paradigm for enabling in-flow, context-aware learning of emerging software engineering skills.

Looking ahead, we plan to conduct larger-scale studies with longer engagement durations and additional measures such as retention and downstream code quality. Comparative evaluations against traditional learning approaches and other GenAI-based tutors will help isolate the specific impact of PC. We also aim to deploy PC in real-world project settings to evaluate the benefits of context-aware reasoning.  

\section*{Data Availability Statement}
The PC implementation and study data cannot be publicly released due to organizational IP and confidentiality constraints. All aggregate results are reported in full within the paper. Interested members may contact the authors for further discussion.

\bibliographystyle{ACM-Reference-Format}
\bibliography{Bibliography}

\end{document}